%% file: main.tex
\pdfoutput=1
\documentclass{article}
\usepackage{cite}
\usepackage{spconf,amsmath,graphicx}
\usepackage{algorithm}
\usepackage{algorithmic}
\usepackage{hyperref}

\title{Adversarial defense for automatic speaker verification \\ by cascaded self-supervised learning models}

\name{Haibin Wu$^{1^{*}}$, 
      Xu Li$^{2^{*}}$, 
      Andy T. Liu$^1$, 
      Zhiyong Wu$^3$, 
      Helen Meng$^2$, 
      Hung-yi Lee$^1$
\thanks{$*$ Equal contribution.}
}

\address{
$^1$ Graduate Institute of Communication Engineering, National Taiwan University \\
$^2$Human-Computer Communications Laboratory, The Chinese University of Hong Kong \\ 
$^3$Shenzhen International Graduate School, Tsinghua University \\
}
 
\begin{document}
\ninept
\maketitle
\input{0-abstract-keywords}
\vspace{-0.2cm}
\input{1-introduction}
\vspace{-0.2cm}
\input{2-method}
\vspace{-0.2cm}
\input{3-experimental-setup}
\vspace{-0.2cm}
\input{4-experimental-results}
\vspace{-0.5cm}
\input{5-conclusion}
\vspace{-0.2cm}
\input{6-acknowledge}

\bibliographystyle{IEEEbib}
\bibliography{strings,refs}

\end{document}

%% file: 0-abstract-keywords.tex
\begin{abstract}
Automatic speaker verification (ASV) is one of the core technologies in biometric identification.
With the ubiquitous usage of ASV systems in safety-critical applications, more and more malicious attackers attempt to launch adversarial attacks at ASV systems.
In the midst of the arms race between attack and defense in ASV, how to effectively improve the robustness of ASV against adversarial attacks remains an open question.
We note that the self-supervised learning models possess the ability to mitigate superficial perturbations in the input after pretraining.
Hence, with the goal of effective defense in ASV against adversarial attacks, we propose a standard and attack-agnostic method based on cascaded self-supervised learning models to purify the adversarial perturbations.
Experimental results demonstrate that the proposed method achieves effective defense performance and can successfully counter adversarial attacks in scenarios where attackers may either be aware or unaware of the self-supervised learning models.
\end{abstract}

%% file: 1-introduction.tex
\section{Introduction}
\label{sec:intro}
Automatic speaker verification (ASV) aims at confirming a speaker identity claim given a segment of spoken utterance. 
The technology has been widely applied in our everyday lives, such as smart phones, e-banking authentication, etc.
Through decades of development, three most representative model architectures with high-performance were proposed, i.e. i-vector embedding systems \cite{lei2014novel,kenny2012small,dehak2010front,garcia2011analysis}, x-vector embedding systems \cite{snyder2018x,li2020bayesian} and r-vector embedding systems \cite{chung2018voxceleb2,li2018deep}.
ASV is one of the most essential technologies for biometric identification, so the security for ASV systems is vitally important.
However, previous work have shown that cutting-edge ASV systems are not only subjected to spoofing audios \cite{yamagishi2019asvspoof} generated by audio replay, speech synthesis and voice conversion, they are vulnerable to adversarial attacks as well \cite{szegedy2013intriguing, kreuk2018fooling, li2020adversarial, xie2020real,wang2020inaudible,das2020attacker}.

The concept of adversarial attacks was first proposed by Szegedy et al. \cite{szegedy2013intriguing} and they showed that an image classification neural network that can outperform humans on classification of clean testing images can become seriously confused on the same testing set after some imperceptible adversarial perturbations are added.
Adversarial samples are composed of genuine samples and deliberately crafted adversarial perturbations, and using adversarial samples to attack well-trained neural networks is called adversarial attack.
Not only can adversarial perturbations make image classification models fail catastrophically, such attacks can also affect speech-related tasks.
Carlini et al. \cite{carlini2018audio} investigated the vulnerability of end-to-end automatic speech recognition (ASR) models by targeted adversarial attacks.
Given a piece of audio, whether speech or music, they can craft another adversarial audio, that is over 99\% similar to the original one, but can manipulate the ASR model to hallucinate arbitrarily predefined transcriptions.
The anti-spoofing model, a protector for ASV systems by detecting and filtering spoofing audios, can also be subjected to adversarial attacks \cite{liu2019adversarial}. 
This was among the first efforts to show that high-performance anti-spoofing models cannot counter adversarial attacks in both white-box and black-box scenarios.

With the ubiquitous usage of ASV systems in safety-critical environments, more and more malicious attackers attempt to launch adversarial attacks at ASV systems \cite{kreuk2018fooling, li2020adversarial, xie2020real,wang2020inaudible}.
\cite{kreuk2018fooling} first adopted adversarial samples to deceive the end-to-end ASV systems. 
They conducted both cross-dataset and cross-feature attacks and showed the effectiveness of adversarial samples in both settings.
Even the state-of-the-art ASV models, GMM i-vector system and x-vector system, are vulnerable to adversarial attacks \cite{li2020adversarial}.
Also, \cite{li2020adversarial} illustrated the adversarial samples generated from i-vector systems are transferable to attack the x-vector systems.
Xie et al. \cite{xie2020real} crafted more dangerous adversarial samples which were universal, real-time and robust to deceive the x-vector based speaker recognition systems.
\cite{wang2020inaudible} employed the psychoacoustic principle of frequency masking to make the adversarial audios against x-vector based speaker recognition system more indistinguishable to original audios from human's perception.


Adversarial perturbations on ASV models have compromised their robustness considerably, which makes them unreliable in some safety-critical environments. 
This has led researchers to develop a variety of defense methods to counter the attacks. 
Wang et al. \cite{wang2019adversarial} injected the adversarial samples into the training set and adopted the adversarial objective as regularization to improve the robustness of ASV.
Adversarial training which adopts adversarial samples to augment the training set was introduced to alleviate the vulnerability of anti-spoofing for ASV against adversarial attacks \cite{wu2020defense}.
Li et al. \cite{li2020investigating} separately trained a detection network to distinguish the adversarial samples from genuine samples. 
A major drawback of the three methods \cite{wang2019adversarial,wu2020defense,li2020investigating} is that they need to know the details of the attacking algorithm for adversarial sample generation.
So the above three methods tend to overfit to the attacking algorithm used for generating adversarial samples which are used for training the defense models, not to mention that it is impossible for the ASV system designer to know the exact attacking algorithm adopted by the attackers in the wild.
Wu et al. 
\cite{wu2020mock} proposed to use the self-supervised learning based model, the Mockingjay \cite{mockingjay}, as a feature extractor in front of the anti-spoofing of ASV to mitigate the transferability of black-box attacks. This method requires modification and retraining of the anti-spoofing model, and is confined to only black-box attack scenarios.

Self-supervised learning arouses keen interests recently, and transformer encoder representations from alteration (TERA) \cite{tera} is a self-supervised learning method proposed as a more advanced approach to Mockingjay~\cite{mockingjay}.
The TERA model is trained by a denoising task. After training, it possesses the ability of mitigating superficial perturbations in the inputs and transforming corrupted speech into clean speech.
The adversarial perturbations can be considered as a kind of noise and thereafter, the pretrained TERA models can also counter the adversarial noise to some extent.

In the midst of the arms race between attack and defense for ASV, how to improve the robustness of ASV against adversarial perturbations still remains as an open question.
Hence, this work proposes the cascaded TERA models to purify the adversarial perturbations and counter adversarial attacks.
The proposed defense method is a standard method without changing the internals of ASV systems.
So it has no conflict with previous defense methods \cite{wang2019adversarial,wu2020defense,li2020investigating} and can even serve as reinforcement for them.
Also, in contrast to previous attack-dependent methods \cite{wang2019adversarial,wu2020defense,li2020investigating}, the proposed method is an attack-agnostic method which doesn't require the knowledge about adversarial samples generation process.
As the beginning work for adversarial defense of ASV by attack-agnostic methods, there is no baseline for reference.
So we also first employ hand-crafted filters for adversarial defense and set them as our baseline.
Our contributions are as follows:
We are among the first to propose the self-supervised learning based models for adversarial defense on ASV systems.
We begin with applying hand-crafted filters including Gaussian, mean and median filters, to counter adversarial attacks for ASV. 
Experimental results demonstrate that our proposed method achieves effective defense performance and successfully counter adversarial attacks in both scenarios where attackers are aware or unaware of self-supervised learning models.

%% file: 2-method.tex
\section{Proposed method}
\label{sec:method}

\begin{figure*}[ht]
  \centering
  \centerline{\includegraphics[width=\linewidth]{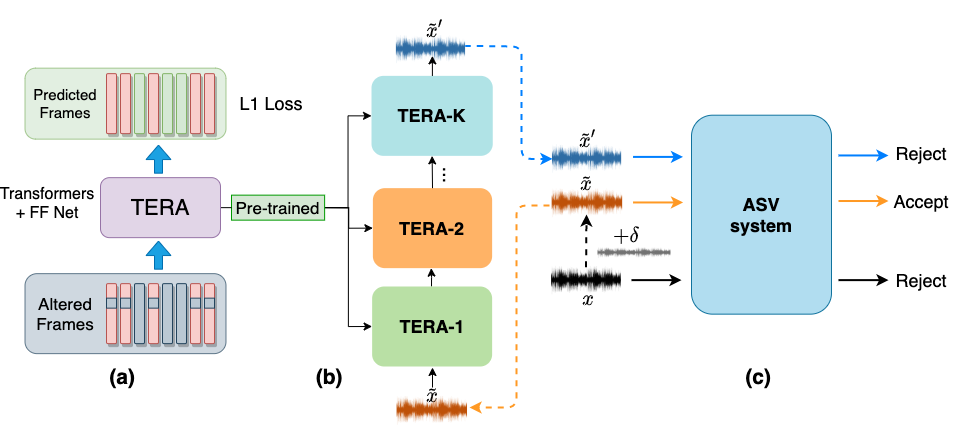}}
  \caption{(a). The illustration of TERA's pretraining strategy. (b) The framework for adversarial defense on ASV by TERA models. (c). The procedure of adversarial attack.}
  \label{fig:cascade of Tera}
\end{figure*}

\subsection{TERA pretraining}
\vspace{-0.15cm} 
The TERA model is pretrained by solving a self-supervised alteration-prediction task with a $L_1$ reconstruction loss function as shown in Fig.~\ref{fig:cascade of Tera}a.
At training time, the TERA pretraining task requires the model to take a sequence of frames as input that has a certain percentage of randomly selected portions to be altered, and attempts to reconstruct the altered frames.
The TERA pretraining scheme consists of several objectives:
1) time alteration: reconstructing from corrupted blocks of time steps with width of $W_T$.
2) channel alteration: reconstructing from missing blocks of frequency channels with width of $W_C$.
3) magnitude alteration: reconstructing from altered feature magnitudes with a probability of $P_N$.
The model acquires information around the corrupted or altered input, and use them to reconstruct the clean input. After pretraining, the model learns the ability to map corrupted speech to clean speech, and also the ability of denoising and purification.

\subsection{Defense method by cascaded TERA models}
\vspace{-0.1cm} 
This subsection will present the defense procedure of our proposed method.
As shown in Fig.~\ref{fig:cascade of Tera}c, the in-the-wild attackers can deliberately find adversarial noise $\delta$ and add it to the genuine sample $x$ to generate the adversarial sample $\Tilde{x}$.
The adversarial sample $\Tilde{x}$ is over 99\% similar to the genuine sample $x$ from human's ears, but the prediction of the ASV model reverses.
The adversarial attacks which make ASV models fail catastrophically are dangerous.
Hence, this paper proposes the cascaded self-supervised learning models to counter them.
Fig.~\ref{fig:cascade of Tera}b illustrates the framework of adversarial defense by integrating ASV with cascaded TERA models. We first pretrain the TERA model, as shown in Fig.~\ref{fig:cascade of Tera}a.
Then we define $K$, the number of concatenated TERA models.
Concatenating $K$ TERA models should reduce the adversarial attack success rate without sacrificing the accuracy of benign samples too much.
We show the procedure of finding such a qualified $K$ in section~\ref{subsec:limited}.
Ideally, given a piece of adversarial audio $\Tilde{x}$, the cascaded TERA models will serve as a deep filter to help decontaminate the superficial adversarial perturbations and reconstruct the pivotal information from the input.
If the input is a piece of genuine audio $x$, the deep filter simply performs nearly lossless reconstruction and keep the key information.
After purification, the purified audio $\Tilde{x}'$ will be used for ASV tasks.

\subsection{Threat model and our countermeasure}
\vspace{-0.15cm} 
Previous works coarsely divide the attacking scenarios into white-box and black-box, which is vague and ambiguous to some extent.
In contrast, we attempt to detail the attacking scenario and name two threat models from the perspective of the adversary's knowledge.
Thereafter, we will present our countermeasures.
\begin{itemize}
    \item \textit{Adversaries are unaware of TERA}: Attackers have the access to the internals of the target ASV model, including model structure, model parameters and gradients, while they are not aware of the existence of the cascaded TERA models in front of the ASV model. In this setting, the cascaded TERA models serve as a deep filter to decontaminate the adversarial samples. The TERA obtains the ability of purifying corrupted speech into clean speech after pretraining. So as we will see in subsection~\ref{subsec:limited}, when the number of cascaded TERA models increases, the equal error rate decreases, which shows the effectiveness of the proposed method in mitigating adversarial attack against these adversaries.
    \item \textit{Adversaries are aware of TERA}: Attackers have access to the entire ASV model and the training strategy of TERA models. Our experiments show that even though attackers generate adversarial samples with some information of the TERA models, our approach is still effective on purifying adversarial samples and protecting the ASV models.
\end{itemize}

%% file: 3-experimental-setup.tex
\section{Experimental setup}
\label{sec:expt-setup}

\subsection{ASV setting}
\vspace{-0.15cm} 
This work adopts the r-vector embedding system \cite{zeinali2019but} as the ASV to be attacked. The notation of r-vector comes from the network architecture of ResNet \cite{zeinali2019but}, which has been adopted in the state-of-the-art speaker verification systems. The r-vector system adopts the same architecture as \cite{zeinali2019but}, and AAM-softmax loss \cite{xiang2019margin} with hyper-parameters \{m = 0.2, s = 30\} is used for training neural networks. Extracted r-vectors are length-normalized before cosine scoring.

\subsection{Adversarial samples generation}
\vspace{-0.15cm} 
In this work, we generate adversarial samples using the basic iterative method (BIM) \cite{kurakin2016adversarial} to attack the r-vector system, which has been verifed to be effective to degrade deep neural network systems. We assume that $\boldsymbol{X^{(e)}}$ and $\boldsymbol{X^{(t)}}$ are enrollment and testing utterances, respectively. The ASV system function is denoted as $S$ with parameters $\boldsymbol{\theta}$. 
Attackers aims at perturbing the genuine testing input $\boldsymbol{X^{(t)}}$ to make it more similar with $\boldsymbol{X^{(e)}}$ under the judgement of ASV.
By applying BIM, it perturbs $\boldsymbol{X^{(t)}}$ towards the gradient of system output $S$ w.r.t. $\boldsymbol{X^{(t)}}$ in an iterative manner. Starting from the genuine input $\boldsymbol{X^{(t)}_0} = \boldsymbol{X^{(t)}}$, this process can be formulated as Eq.~\ref{eq:bim-solu}:
\begin{align}
   \nonumber
    \boldsymbol{X^{(t)}_{n+1}} = clip_{\boldsymbol{X^{(t)}}, \epsilon}(\boldsymbol{X^{(t)}_{n}}+\alpha si&gn(\nabla_{\boldsymbol{X^{(t)}_{n}}} S_{\boldsymbol{\theta}}(\boldsymbol{X^{(e)}}, \boldsymbol{X^{(t)}_{n}}))), \\
    &\text{for $n = 0, ..., N-1$} \label{eq:bim-solu}
\end{align}
where $sign$ is a function that takes the sign of the gradient, $\alpha$ is the step size, $N$ is the number of iterations, $\epsilon$ is the perturbation degree and $clip_{\boldsymbol{X^{(t)}}, \epsilon}(\boldsymbol{X})$ holds the norm constraints by applying element-wise clipping such that $\Vert \boldsymbol{X}-\boldsymbol{X^{(t)}} \Vert_{\infty} \leq \epsilon$. In our experiments, $N$ is set as 5. The $\epsilon$ is set as 0.3, so that there is no difference between adversarial and genuine audios from human perception, while the attack can still succeed in making the ASV system behave incorrectly. Finally, $\alpha$ is set as $\epsilon$ divided by $N$.

\subsection{Dataset}
\vspace{-0.15cm} 
This work is conducted on Voxceleb1 \cite{nagrani2017voxceleb}, which consists of short clips of human speech. There are in total 148,642 utterances for 1251 speakers. We develop our ASV system on the training and development partitions, while reserve 4,874 utterances of the testing partition for evaluating our ASV system and generating adversarial samples.
Notice that generating adversarial samples is time-consuming and resource-consuming. Without loss of generality, we randomly select 1000 trials out of 37,720 trials provided in \cite{nagrani2017voxceleb}, to generate adversarial samples.

\subsection{ASV performance with genuine and adversarial inputs}
\vspace{-0.15cm} 
Table~\ref{tab:r-vec-performance} shows the r-vector system performance on the complete trials provided in \cite{nagrani2017voxceleb} and also the selected 1K trials. The system performance is evaluated by equal error rate (EER) and minimum detection cost function (minDCF) with a prior probability of target trials to be 0.01. We observe that the system performance has been seriously degraded after applying adversarial inputs, which verifies the effectiveness of our adversarial attack algorithm. Besides, we observe consistent performance trends between the results on the complete trials and those on the selected 1K trials, which indicates that the selection process of 1K trials is reasonable. Further experiments are conducted on these 1K adversarial samples.

\begin{table}[t]
\vspace{-0.4cm} 
\caption{The r-vector system performance with genuine (gen-input) and adversarial inputs (adv-input).}
\label{tab:r-vec-performance}
\centering
\begin{tabular}{c|c|c|c|c}
\hline
\hline
 & \multicolumn{2}{c}{complete trials} & \multicolumn{2}{|c}{1K trials} \\
 \cline{2-5}
            & EER (\%) & minDCF & EER (\%) & minDCF \\
\hline
gen-input     & 8.39    & 0.638       & 8.87 & 0.792 \\
adv-input    &  65.92    & 1.000        & 66.02 & 1.000 \\
\hline
\hline
\end{tabular}
\end{table}

\subsection{The setting of TERA's pretraining}
\vspace{-0.15cm} 
We use the TERA implementation from the S3PRL speech toolkit. 
We use a time alteration width $W_T$ of $7$, channel alteration width $W_C$ of $5$, and magnitude alteration probability $P_N$ set to $0$.
$7$ frames of time alteration corresponding to $70$ ms of speech, which is in the range of average phoneme duration.
According to \cite{park2019specaugment}, we set $W_C$ as $5$ for better reconstruction.
Setting $W_C$ and $W_T$ too large will make the self-supervised learning model hard to do reconstruction.
The rest of the alteration policy follows the original design of TERA~\cite{tera}.
In order to evaluate our proposed method in the scenario where adversaries are aware of TERA, we pretrain two TERA models with an identical setting except for a unique random seed, denoted as TERA0 and TERA1, respectively.
Each model consists of 3 layers of Transformer encoders with multi-head self-attention~\cite{vaswani2017attention}, followed by a feed-forward prediction network.
The dataset adopted to pretrain TERA models is Voxceleb2 \cite{chung2018voxceleb2}.
The input to the model is 24-dim MFCC extracted by standard Kaldi~\cite{kaldi} scripts.
We use the Adam optimizer \cite{kingma2014adam} with mini-batches of size 128 to find model parameters that minimize the L1 loss of the TERA pretraining task.
The model is trained for 30K steps, where learning rate is warmed up over the first 7\% to a peak value of $2 \times 10^{-4}$ and then linearly decayed.
If not specified otherwise, other pretraining settings follow the TERA paper~\cite{tera}.

%% file: 4-experimental-results.tex
\section{Experimental Results}
\label{sec:expt-rst}

\subsection{Adversaries are unaware of TERA}
\label{subsec:limited}
\begin{figure}[ht]
\vspace{-0.3cm} 
    \centering
    \includegraphics[width=0.45\textwidth]{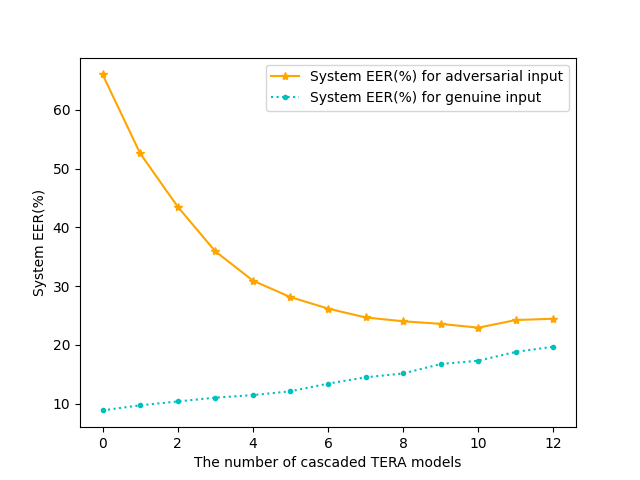}
    \caption{The r-vector system's EER (\%) for ASV integrated with different number of cascaded TERA models.}
    \label{fig:sys_performance_with_tera_number}
\end{figure}

\vspace{-0.15cm} 
This subsection assumes that attackers have access to the complete ASV model parameters while they are unaware of the cascaded TERA models in the frontend of the ASV system.
This setting is the most practical one in the real world because it is unrealistic for attackers to know everything about the target models through querying the API.
In this setting, we adopt TERA0 as a basic element, and duplicate it into different amounts to be placed in front of the ASV. 
Defense performance is evaluated when ASV integrated with different number of TERA models, as shown in Fig.~\ref{fig:sys_performance_with_tera_number}.

We observe that for adversarial speech, integration of TERA models can dramatically decrease EER of the attacked system from over 65\% to around 20\%, which indicates that integration of TERA models can purify the adversarial signals and mitigate the attack effectiveness.
We also observe that the performance of ASV with genuine inputs drops due to the imperfect reconstruction of TERA models. Possible solutions can either be improving the reconstruction ability of TERA, or using reconstructed inputs to finetune ASV systems, which will be investigated in future works.

As investigated in \cite{wu2020defense}, some hand-crafted filters also have the ability of purifying adversarial signals and alleviating the destructiveness of adversarial attacks. In this work, we leverage three filters, i.e. Gaussian, median and mean filters, to be positioned in front of the ASV to defend against adversarial attacks and set them as our baseline. Table~\ref{tab:defense-performance-tera-filters} illustrates the system EER for genuine and adversarial inputs when the ASV integrated with cascaded TERA models, Gaussian, median and mean filters. We observe that all filters have the ability of purifying adversarial signals given adversarial inputs. The attack effectiveness has been degraded by over 50\% after integration with these filters. However, due to additional noise caused by the filtering process, all filters also degrade the system performance on genuine inputs. Notice that the proposed TERA models outperform the other filters with respect to both purifying adversarial signals within adversarial inputs, and preserving ASV performance on genuine inputs.
\subsection{Adversaries are aware of TERA}
\label{subsec:perfect-knowledge-adversary}
\vspace{-0.15cm} 
This subsection gives a case study to show the robustness of our defense approach to a more severe attacking scenario, where attackers not only have access to the entire ASV parameters, but also are aware of the TERA models in front of ASV. We assume that attackers know the training strategy of the TERA model, and pretrain a substitute TERA model (denoted as TERA1) to be placed in front of ASV to generate adversarial samples. In the real-world applications, it is hard for attackers to know the specific number of TERA models in front of ASV, and in this work we only integrate ASV with one TERA1 model to generate adversarial samples. The attacking process is identically configured as BIM with perturbation degree $\epsilon=0.3$. Table~\ref{tab:eer-aware-of-TERA} illustrates the system performance when ASV integrated with different number of TERA0 models, given adversarial samples generated by the integration of ASV and one TERA1 model as the inputs. 
We observe that even though attackers know the training setting of TERA models in front of ASV, the attack destructiveness is still alleviated by integrating ASV with more TERA models. Moreover, based on our experiments, it is memory- and computation-consuming when performing white-box attacks on ASV integrated with TERA models. 
Hence placing sufficient number of TERA models in front of ASV could be a good option to defend against adversarial attacks.

\begin{table}[t]
\vspace{-0.3cm} 
\caption{The system's EER(\%) for genuine and adversarial inputs when integrating ASV with TERA, Gaussian, median and mean filters. (NA means nothing is positioned in front of ASV.)}
\label{tab:defense-performance-tera-filters}
\centering
\begin{tabular}{c|c|c|c|c|c}
\hline
\hline
        & NA & 10*TERA0 & Gaussian & median & mean \\
\hline
gen-input    & 8.87 & 17.32    & 30.30 & 27.06 & 27.71 \\
adv-input & 66.02 & 22.94   &  31.60 & 29.65 & 29.44 \\
\hline
\hline
\end{tabular}
\vspace{-0.3cm}
\end{table}


\begin{table}[h]
\vspace{-0.3cm} 
\caption{The r-vector system's EER(\%) when integrating ASV with different number of TERA models.}
\label{tab:eer-aware-of-TERA}
\centering
\begin{tabular}{c|c|c|c|c}
\hline
\hline
         & NA & 1*TERA0 & 2*TERA0 & 3*TERA0 \\
\hline
 EER(\%) & 54.55    & 53.68 & 47.62   &   40.69 \\
\hline
\hline
\end{tabular}
\vspace{-0.3cm} 
\end{table}

%% file: 5-conclusion.tex
\section{conclusion}
\label{sec:conclusion}
This work proposes integrating ASV with cascaded TERA models for defense against adversarial attacks.
We conduct experiments in two attacking scenarios depending on whether the adversaries are aware of the TERA models or not.
The scenario where attackers are unaware of the TERA models is more practical, and experimental results indicate that introducing cascaded TERA models as a deep filter can purify the adversarial signals and mitigate the attack destructiveness.
Also our proposed method outperforms hand-crafted filters with respect to both decontaminating adversarial signals within adversarial inputs, and preserving ASV performance on genuine inputs.
For the other scenario where attackers are aware of the TERA models, experimental results verify that integrating ASV with more TERA models is still effective on alleviating adversarial noise even though attackers utilize the TERA information to generate adversarial samples.
Preserving better performance on genuine samples than hand-crafted filters is not good enough, so we attempt to tackle this problem in future works.

%% file: 6-acknowledge.tex
\section{acknowledgement}
\label{sec:acknowledge}
This work was done when H. Wu was a visiting student at Shenzhen International Graduate School, Tsinghua University.
H. Wu and A. Liu are supported by Frontier Speech Technology Scholarship of National Taiwan University. A. Liu is supported by ASUS AICS.
Xu Li is supported by HKSAR Government's Research Grants Council General Research Fund (Project No. 14208718).